# The MAGNEX magnetic spectrometer for double charge exchange reactions

M. Cavallaro[1], C. Agodi[1], G. Brischetto[1,2], S. Calabrese[1,2], F. Cappuzzello[1,2], D. Carbone[1], I. Ciraldo[1,2], A. Pakou[3], O. Sgouros[1], V. Soukeras[1], G. Souliotis[4], A. Spatafora[1,2], D. Torresi[1]

for the NUMEN collaboration

[1] Istituto Nazionale di Fisica Nucleare, Laboratori Nazionali del Sud, Catania, Italy

[2] Dipartimento di Fisica e Astronomia, Università di Catania, Catania, Italy

[3] Department of Physics and HINP, The University of Ioannina, Ioannina, Greece

[4] Laboratory of Physical Chemistry, Department of Chemistry, National and Kapodistrian University of Athens, Athens

**Abstract**

Physics cases of increasing interest in the recent years, such as the study of double charge exchange reactions for neutrino physics, require the study of very suppressed reaction channels in medium-heavy ion induced nuclear reactions. The main experimental challenges are the possibility to identify and detect the medium-heavy quasi-projectiles and the capability to measure very low cross-sections (few nb) with high sensitivity at very forward angles, including zero degree. The experimental techniques adopted in the setup of the MAGNEX magnetic spectrometer to face these issues are described in this paper.

## 1. Introduction

An innovative technique to access the nuclear matrix elements entering the expression of the lifetime of the double beta decay by cross sections measurements of heavy-ion induced Double Charge Exchange (DCE) reactions has been recently proposed at INFN-LNS within the NUMEN [1], [2] and NURE projects [3]. Despite the double beta decay and double charge exchange processes are different as they are triggered by the weak and strong interaction respectively, important similarities are present. The basic ones are the coincidence of the initial and final nuclear state wave-functions and the formal similarity of the transition operators, which in both cases present a superposition of short range Fermi, Gamow-Teller and rank-two tensor components with a sizable momentum available in the transition. First pioneering experimental results obtained at the INFN-LNS laboratory for the $^{40}$Ca($^{18}$O,$^{18}$Ne)$^{40}$Ar reaction at 270 MeV have given encouraging indication on the capability of the proposed technique to access relevant quantitative information [4].

From the experimental point of view, key aspects of the project, in addition to the availability of good quality medium-heavy ion beams, are:

(i) the possibility to identify and detect the medium-heavy quasi-projectiles;

(ii) the possibility to get high energy and angular resolution to distinguish transitions to individual states and measure accurate differential cross-sections;

(iii) the capability to measure very low cross-sections (few nb) with high sensitivity at very forward angles, including zero degree.

Such experimental challenges are faced at INFN-LNS by using the K800 Superconducting Cyclotron to accelerate beams and the MAGNEX large acceptance magnetic spectrometer for the detection of the ejectiles. In this paper, the features related to MAGNEX are discussed. The main characteristics of the spectrometer are described in Section 2, the experimental techniques for the zero-degree measurement are presented in Section 3 and the particle identification technique, upgraded for the discrimination of different charge states by the introduction of the time of flight measurement, is illustrated in Section 4.

## 2. The MAGNEX spectrometer

The MAGNEX spectrometer is a large acceptance magnetic device made up of a large aperture vertically focusing quadrupole and a horizontally bending dipole magnet [5].

MAGNEX was designed to investigate processes characterized by very low yields and allows the identification of heavy ions with quite high mass ($\Delta A/A \sim 1/160$), angle ($\Delta\theta \sim 0.2°$) and energy resolutions ($\Delta E/E \sim 1/1000$), within a large solid angle ($\Omega \sim 50$ msr) and momentum range ($-14\% < \Delta p/p < +10\%$). High-resolution measurements for quasi-elastic processes, characterized by differential cross-sections falling down to tens of nb/sr, were already performed by this setup [6], [7], [8]. A crucial feature is the implementation of the powerful technique of trajectory reconstruction, based on differential algebraic techniques, which allows solving the equation of motion of each detected particle to 10$^{th}$ order [9], [10], [11], [12], [13]. This is a unique characteristic of MAGNEX, which guarantees the above mentioned performances and its relevance in the worldwide scenario of heavy-ion physics [14], [15], [16], [17], [18] also taking advantage of its coupling to the EDEN neutron detector array [19].

The Focal Plane Detector (FPD) consists of a large (active volume 1360 mm x 200 mm x 96 mm) low-pressure gas-filled tracker followed by a wall of 60 silicon pad detectors to stop the particles. A set of wire-based drift chambers measures the vertical position and angle of the reaction ejectiles, while the induced charge distributions on a set of segmented pads allow to extract the horizontal position and angle [20]. The energy loss measured by the multiplication wires and the residual energy at the silicon detectors are used for atomic number identification of the ions. The $\sqrt{m}/q$ identification is performed exploiting the relation between measured position and energy in the dispersive direction, as described in [21] and in Section 4. The present FPD is a suitable detector to discriminate from light to heavy ions with 0.6% resolution in $\sqrt{m}/q$ and 2% in atomic number. The tracking measurement sensitivity guarantees an overall energy resolution of about 1/1000, which is close to the limit of the optics for the used beams. The performances of the present FPD are described in ref. [22].

### 3. The zero-degree measurement

The NUMEN and NURE experimental activity with accelerated beams consists of two main classes of experiments, corresponding to the exploration of the two directions of isospin lowering $\tau^-\tau^-$ and rising $\tau^+\tau^+$, characteristic of $\beta^-\beta^-$ and $\beta^+\beta^+$ decays respectively.

In particular, the $\beta^+\beta^+$ direction in the target is investigated using an $^{18}O^{8+}$ beam and measuring the ($^{18}O$,$^{18}Ne$) Double Charge Exchange (DCE) induced transitions, together with other reaction channels involving same beam and target. Similarly, the $\beta^-\beta^-$ direction is explored via the ($^{20}Ne$,$^{20}O$) reaction, using a $^{20}Ne^{10+}$ beam and detecting the reaction products of the DCE channel and of the other open channels characterized by same projectile and target.

Exploratory investigations of the two classes of experiments have been already performed, highlighting the strengths and the limiting aspects of the adopted technique and establishing the best working conditions [23].

*3.1 Experiments with $^{18}O$ beam ($\beta^+\beta^+$ direction)*

For the experiments of this class, reaction channels of our interest are listed below:
- Elastic and inelastic scattering ($^{18}O$,$^{18}O$)
- DCE reaction ($^{18}O$,$^{18}Ne$)
- Charge-exchange reaction ($^{18}O$,$^{18}F$)
- Two-proton pickup reaction ($^{18}O$,$^{20}Ne$)
- One-proton pickup reaction ($^{18}O$,$^{19}F$)
- Two-neutron stripping reaction ($^{18}O$, $^{16}O$)
- One-neutron stripping reaction ($^{18}O$,$^{17}O$)

The purpose is to build a coherent set of experimentally driven information on nucleus-nucleus potentials and projectile/target wave functions, thus providing stringent constraints to the theoretical calculations.

One of the main challenges of such experiments is the measurement at very forward angles, including zero-degree. This is performed by placing the spectrometer with its optical axis at +3° with respect to the beam axis. Thanks to its large angular acceptance, a range -2° < $\theta_{lab}$ < 9° is thus covered. The MAGNEX quadrupole and dipole magnetic fields are set in order that the incident beam, after passing through the magnets, reaches a region which is besides the FPD but external to it. For this class of experiments, the incident beam ($^{18}O^{8+}$) has higher magnetic rigidity ($B\rho$) than the ones of the ejectiles of interest (namely $^{18}$Ne, $^{18}$F, $^{20}$Ne, $^{19}$F, $^{16}$O, $^{17}$O with maximum charge state). After passing through the spectrometer, the beam stops in a specifically designed Faraday Cup, which measures the incident charge in each run, placed in the high-$B\rho$ region besides the FPD. The typical beam trajectory and the Faraday cup for this category of experiment are schematically drawn in Fig. 1 as the red line and the red rectangle, respectively. A full high-order simulation, including the tracking of the ion trajectories inside the magnetic field and the complete geometry of the spectrometer, is performed to describe, in each experiment, the actual motion of the beam coming out of the target and of the emitted ejectiles along the spectrometer towards the FPD region.

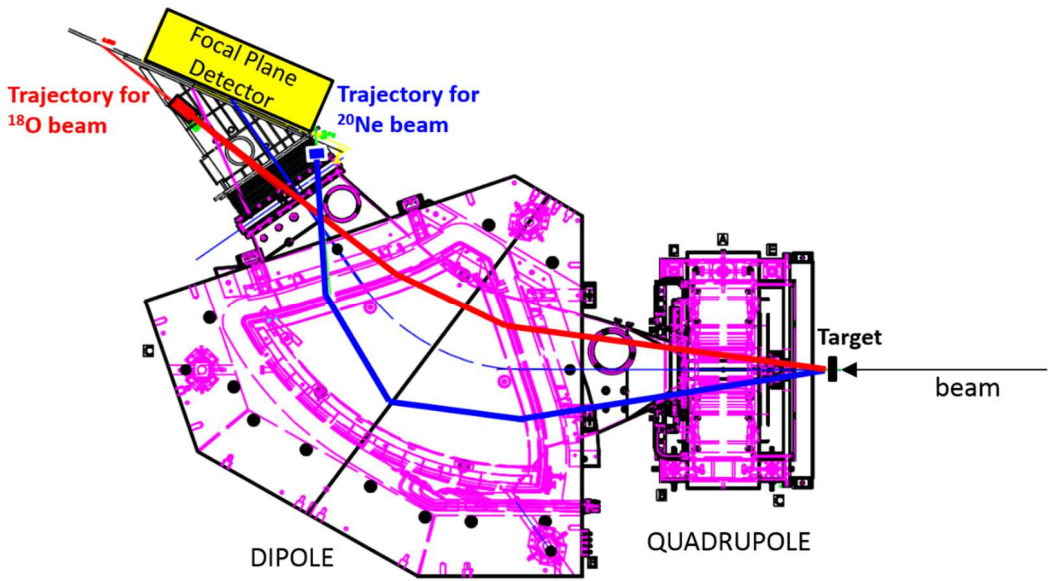

Fig. 1. Layout of the MAGNEX spectrometer. The blue and red lines represent the typical trajectories of the $^{20}Ne^{10+}$ and $^{18}O^{8+}$ beams traversing the spectrometer and reaching the Faraday cups adjacent to the Focal Plane Detector acceptance.

*3.2 Experiments with $^{20}$Ne beam (β⁻β⁻ direction)*

In the class of experiments with $^{20}Ne^{10+}$ beams, the reaction channels we are interested are the following:

- Elastic and inelastic scattering ($^{20}$Ne,$^{20}$Ne)

- DCE reaction ($^{20}$Ne,$^{20}$O)
- Charge Exchange reaction ($^{20}$Ne,$^{20}$F)
- Two-proton stripping reaction ($^{20}$Ne,$^{18}$O)
- One-proton stripping reaction ($^{20}$Ne,$^{19}$F)
- Two-neutron pickup reaction ($^{20}$Ne,$^{22}$Ne)
- One-neutron pickup reaction ($^{20}$Ne,$^{21}$Ne)

For these experiments, the incident beam ($^{20}$Ne$^{10+}$) has a magnetic rigidity which is lower than the reaction ejectiles of interest. Thus, for a fixed magnetic field setting, the beam will be more bent than the ejectiles of interest. The spectrometer optical axis is typically placed at -3°, thus the covered angular range is -8° < $\theta_{lab}$ < +3°. The quadrupole and dipole magnetic fields of MAGNEX are set in order that the $^{20}$Ne$^{10+}$ beam reaches the low-$B\rho$ region besides the FPD as shown in Fig. 1 by the green line (low-$B\rho$ region). A Faraday cup is mounted and properly aligned at this place (green rectangle).

A peculiarity of these experiments concerns the treatment of the different charge states of the beam emerging out of the target. The beam components characterized by charge states lower than 10$^+$, mainly $^{20}$Ne$^{9+}$ and $^{20}$Ne$^{8+}$, produced by the interaction of the beam with the electrons of the target material, have a magnetic rigidity which is similar to the ions of interest. Therefore, they enter in the FPD acceptance, causing a limitation in the rate tolerable by the detector. Such low charge state components of the main beam have in fact an intensity of the order of 10$^{-3}$ (for the 9$^+$) and 10$^{-5}$ (for the 8$^+$), with respect to the 10$^+$ beam. For example, for a typical beam current of 10 enA and at 15 MeV/u beam energy, the amount of 9$^+$ and 8$^+$ components at the focal plane is of the order of 10$^7$ and 10$^5$ pps, respectively. This is beyond the acceptable rate of the FPD. In addition, the elastic scattering on the target at forward angles by $^{20}$Ne$^{9+}$ and $^{20}$Ne$^{8+}$ beams also produces high counting rate at the focal plane.

In order to stop these unwanted $^{20}$Ne particles, two aluminum shields are mounted upstream the sensitive region of the focal plane detector (see Fig. 2). The shields act on a limited phase space region which stops the 9$^+$ and 8$^+$ beams and elastic scattering at very forward angles, but not the other reaction channels generated by these beams. It is known that the charge state distribution of a heavy-ion beam after crossing a material depends on the bombarding energy and on the chemical composition of the target. The targets relevant for NUMEN generate an unwanted charge distribution that can be conveniently changed, minimizing the amount of $^{20}$Ne$^{9+}$ and $^{20}$Ne$^{8+}$, by adding an appropriate second foil (post-stripper) downstream of the isotopic target. Recently a specific study of different materials to be used as post-stripper has been performed [24]. A preliminary analysis of specifically collected data show that material containing Carbon and Hydrogen atoms are the most efficient to reduce the lower charge state contributions and so the most promising in this view. The analysis of these data is presently on going.

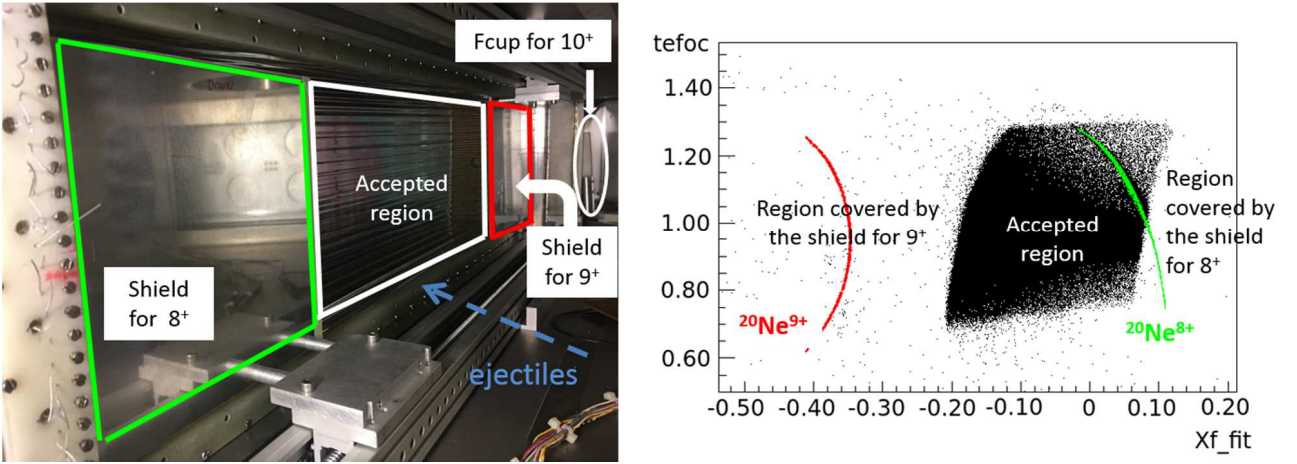

Fig. 2 (left) Picture of the focal plane region of the MAGNEX spectrometer showing the two shields and the Faraday cup used for the $^{20}$Ne$^{10+}$ induced reactions. (right) Plot of the horizontal angle (tefoc) in rad units versus the horizontal position (Xf_fit) in m units measured at the focal plane. The black points are the unidentified experimental data, the red and green lines are the simulated data for the $^{20}$Ne$^{9+}$ and $^{20}$Ne$^{8+}$ beams, respectively.

4. **The particle identification technique**

The ejectiles of interest are typically in the mass number region $18 \leq A \leq 22$ and atomic number region $8 \leq Z \leq 10$. In the usually adopted particle identification technique, the atomic number of the ejectiles is identified by standard $\Delta E$-$E$ correlation plots. A typical $\Delta E$-$E$ two-dimensional plot is shown in Fig. 3 for a single silicon detector together with a graphical contour that includes the Neon ejectiles. The plotted parameters are the residual energy measured by the silicon detectors ($E_{resid}$) and the total energy loss in the FPD gas section $\Delta E^{TOT}_{cor}$ corrected for the different path lengths in the gas.

For the mass identification, an innovative particle identification technique for large acceptance spectrometers, based on the properties of the Lorentz force, was introduced in Ref. [21]. It is based on the relationship between kinetic energy and position ($X_{foc}$) at the focal plane of a bending magnetic element. In the case of MAGNEX it has been demonstrated that it can be simplified using a correlation between $X_{foc}$ and the residual energy measured at the silicon detector $E_{resid}$. The relationship between the two measured quantities ($X_{foc}$ and $E_{resid}$) is approximately quadratic with a factor depending on the ratio $\sqrt{m}/q$

$$X_{foc} \propto \frac{\sqrt{m}}{q}\sqrt{E_{resid}}$$

Therefore, in a $X_{foc}$ vs $E_{resid}$ plot the ions are distributed on different loci according to the ratio $\sqrt{m}/q$. The clear separation between the different Neon isotopes is evident in Fig. 4, where the $X_{foc}$-$E_{resid}$ plot is shown for the data selected with the graphical condition on the $\Delta E^{TOT}_{cor}$-$E_{resid}$ one of Fig. 3. In this representation, the selection of the $^{22}$Ne$^{10+}$ ejectiles can be in principle applied as depicted in Fig. 4 by the red graphical cut.

However, in some cases, the $X_{foc}$-$E_{resid}$ representation of Fig. 4 is not enough to separate ions characterized by different masses and different charge states but with same or similar $\sqrt{m}/q$ ratio. It is the case of the $^{22}$Ne$^{10+}$ and $^{18}$Ne$^{9+}$ ions, which practically lies in the same locus in the $X_{foc}$-$E_{resid}$ plot ($\sqrt{m}/q = 0.469$ for $^{22}$Ne$^{10+}$ and $\sqrt{m}/q = 0.471$ for $^{18}$Ne$^{9+}$).

To face this problem, an improvement of the particle identification technique has been introduced and is presented here for the first time. The time difference between the superconductive cyclotron radiofrequency (RF) periodic signal (STOP) and the timing signal coming from the silicon detectors of the MAGNEX FPD (START) is sent to a time to amplitude converter (TAC) and recorded. The calibrated TAC parameter measured for the events entering in the gate shown in the $X_{foc}$-$E_{resid}$ plot of Fig. 4 clearly allows to separate between the events relative to the $^{22}$Ne$^{10+}$ and the $^{18}$Ne$^{9+}$ ejectiles.

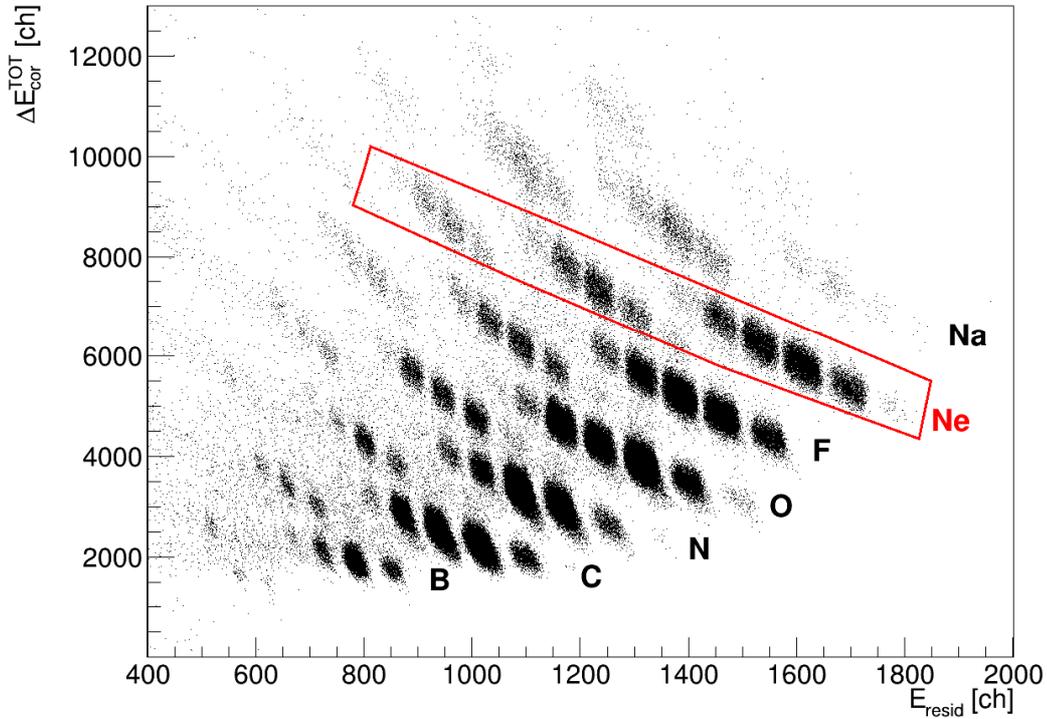

Fig. 3. Typical $\Delta E^{TOT}_{cor}$ vs $E_{resid}$ plot for a single silicon detector. The different atomic species and a graphical contour on the Neon region are indicated.

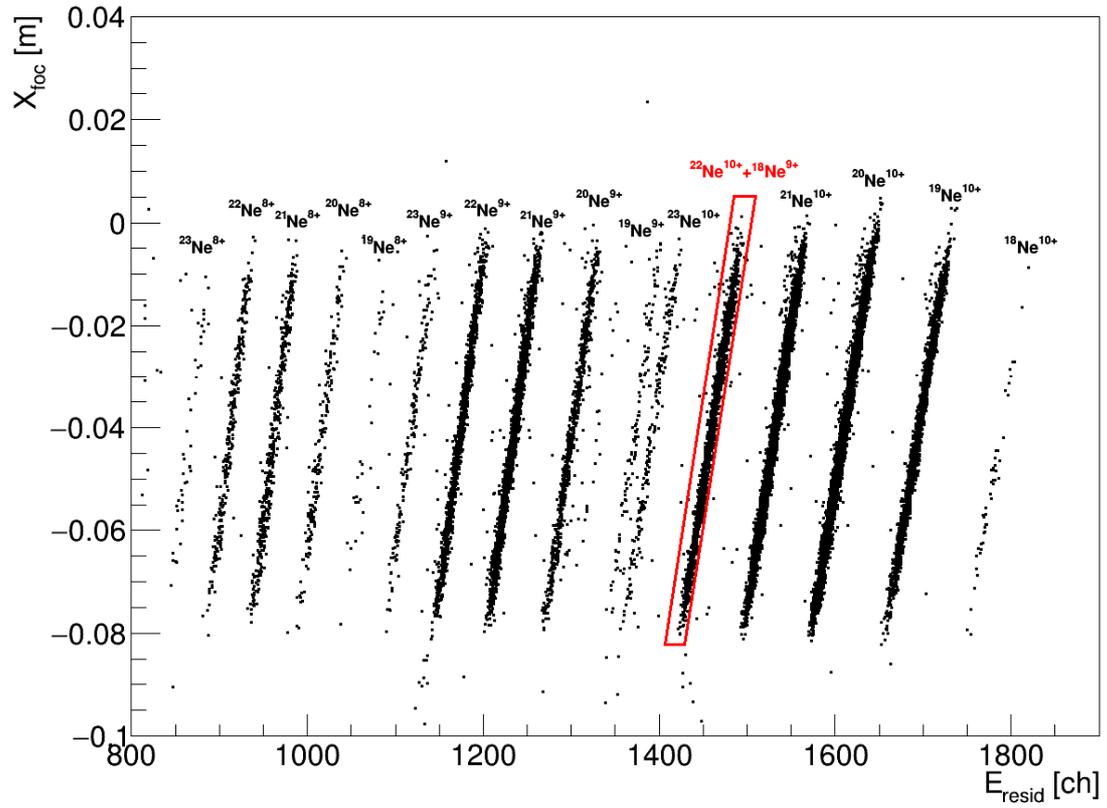

Fig. 4. Typical $X_{foc}$-$E_{resid}$ plot after applying the graphical condition shown in Fig. 3 on the $\Delta E^{TOT}_{cor}$−$E_{resid}$ for the same silicon detector. The different Neon isotopes and a graphical contour selecting the $^{22}$Ne$^{10+}$ and the $^{18}$Ne$^{9+}$ ejectiles are indicated.

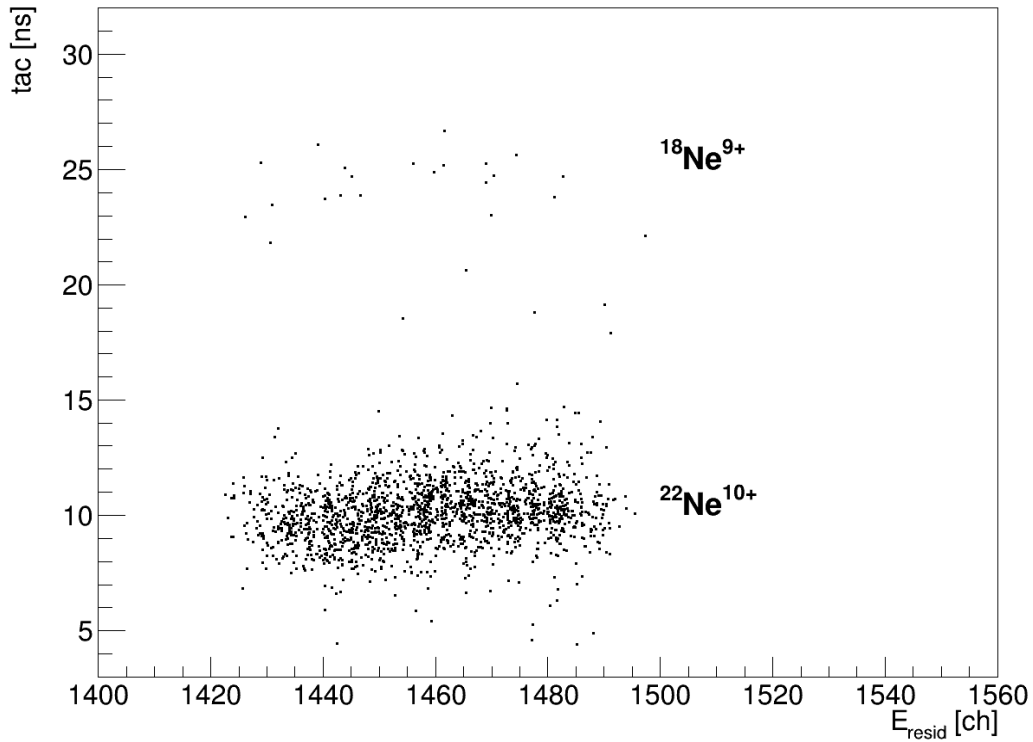

Fig. 5. Typical TAC vs $E_{resid}$ plot for the events gated by the graphical contours shown in Fig. 3 and Fig. 4. The $^{22}Ne^{10+}$ and the $^{18}Ne^{9+}$ loci are indicated.

**Conclusions**

In the present paper, the techniques adopted to set up the MAGNEX spectrometer for the challenging measurements of very suppressed reaction channels in medium-heavy ion induced nuclear reactions, such as the double charge exchange experiments, are described.

The strategy used to perform the zero-degree measurement, based on an accurate simulation of the ion trajectories along the spectrometer and on the use of two Faraday cups properly designed for each category of experiments, is described.

The technique typically used in MAGNEX for particle identification has been upgraded introducing the measurement of the time difference between the superconductive cyclotron RF signal and the timing signal coming from the silicon detectors of the MAGNEX FPD. This technique is described in the present paper for the first time and the experimental results are shown. The possibility to perform high performing particle identification for medium mass ejectiles characterized by different charge states is demonstrated.

**Acknowledgements**

This project has received funding from the European Research Council (ERC) under the European Union's Horizon 2020 research and innovation programme (grant agreement No 714625)